\documentclass{article}

\usepackage{arxiv}
\pdfoutput=1
 
\usepackage[utf8]{inputenc} % allow utf-8 input
\usepackage[T1]{fontenc}    % use 8-bit T1 fonts
\usepackage{hyperref}       % hyperlinks
\usepackage{url}            % simple URL typesetting
\usepackage{booktabs}       % professional-quality tables
\usepackage{amsfonts}       % blackboard math symbols
\usepackage{nicefrac}       % compact symbols for 1/2, etc.
\usepackage{microtype}      % microtypography
\usepackage{lipsum}
\usepackage[right]{lineno}  % line numbers
\usepackage{adjustbox}      % adjust tables
\usepackage[aboveskip=1pt,labelfont=bf,labelsep=period,singlelinecheck=off]{caption} % adjust caption style
\usepackage{microtype}
\DisableLigatures[f]{encoding = *, family = * } % improves typesetting
\usepackage{color}          % for colored text
\definecolor{Gray}{gray}{.25}
\usepackage{cprotect}       % verbatim in caption
\usepackage{multirow}
\usepackage{listings}
\usepackage{natbib,hyperref}
\usepackage{lineno}
\usepackage[dvipsnames]{xcolor}

\definecolor{backcolour}{rgb}{0.97,0.97,0.97}
\lstdefinestyle{mystyle}{
  backgroundcolor=\color{backcolour},   
  basicstyle=\ttfamily\footnotesize,
  breakatwhitespace=false,         
  breaklines=true,                 
  captionpos=b,                    
  keepspaces=true,                 
  showspaces=false,                
  showstringspaces=false,
  showtabs=false,                  
  tabsize=2
}

\author{
  Steffen Ehrmann \orcid{0000-0002-1825-0097} \and
  Ralf Seppelt \orcid{0000-0002-2723-7150} \and
  Carsten Meyer \orcid{0000-0003-3927-5856}
}

\begin{document}
\vspace*{0.35in}

% title goes here:
\begin{flushleft}
{\Large
\textbf\newline{Harmonise and integrate heterogeneous areal data with the R package \texttt{arealDB}}
}
\newline
% authors go here:
\\
\textbf{Steffen Ehrmann}\textsuperscript{1,*}
\textbf{Ralf Seppelt}\textsuperscript{2,3},
\textbf{Carsten Meyer}\textsuperscript{1,3,4*}
\\
\bigskip
\textbf{1} German Centre for Integrative Biodiversity Research (iDiv) Halle-Jena-Leipzig, Deutscher Platz 5e, 04103 Leipzig, Germany
\\
\textbf{2} UFZ – Helmholtz Centre for Environmental Research, Leipzig, Department Computational Landscape Ecology, Permoserstraße 15, 04318 Leipzig, Germany
\\
\textbf{3} Institute of Geosciences and Geography, Martin Luther University Halle-Wittenberg, Halle (Saale), Germany
\\
\textbf{4} Institute of Biology, Leipzig University, Leipzig, Germany
\\
\bigskip
* steffen.ehrmann@idiv.de, carsten.meyer@idiv.de

\end{flushleft}

\begin{abstract}
Many relevant applications in the environmental and socioeconomic sciences use areal data, such as biodiversity checklists, agricultural statistics, or socioeconomic surveys. For applications that surpass the spatial, temporal or thematic scope of any single data source, data must be integrated from several heterogeneous sources. Inconsistent concepts, definitions, or messy data tables make this a tedious and error-prone process. To date, a dedicated tool \textcolor{CornflowerBlue}{to address these challenges} is still lacking.

Here, we introduce the R package \texttt{arealDB} that integrates heterogeneous areal data and associated geometries into a consistent database\textcolor{CornflowerBlue}{, in an easy-to-use workflow}. It is useful for harmonising language and semantics of variables, relating data to geometries, and documenting metadata and provenance. We illustrate the functionality by integrating two disparate datasets (Brazil, USA) on the harvested area of soybean. \texttt{arealDB} promises quality-improvements to downstream scientific, monitoring, and management applications but also substantial time-savings to database collation efforts.
\end{abstract}

% keywords can be removed

\section*{Software availability}
\texttt{arealDB} is an R package that is available \textcolor{CornflowerBlue}{from CRAN via the function \texttt{install.packages(\textquotedbl arealDB\textquotedbl)}}. This paper is based on version v0.3.6, \textcolor{CornflowerBlue}{which is installed via \texttt{devtools::install\_github(\textquotedbl EhrmannS/arealDB@v0.3.6\textquotedbl)}}. All programming was performed by the authors unless stated otherwise.

\section{Introduction}
\label{sec:intro}
Areal data are an essential data type in many socioeconomic applications, from visualising characteristics of human populations, to assessing trade statistics or documenting land ownership.
They are increasingly used to analyse various environmental variables such as global biodiversity patterns.
Areal data are an everyday communication tool in civil society, where they play an important role as illustrative maps in news or education media.
Generally, any phenomenon at the level of finite spatial polygons (called geometries henceforth) \textcolor{CornflowerBlue}{is recorded as areal data}, irrespective of the domain.

Areal data are typically collated and curated by \textcolor{CornflowerBlue}{a diverse set of actors, from} small and focused projects that build datasets \textcolor{CornflowerBlue}{around their particular} observations, for example in nature conservation, to national statistical agencies or intergovernmental organisations such as the World Bank Group\footnote{https://data.worldbank.org/} or the Food and Agriculture Organization\footnote{http://www.fao.org/faostat}.
The data are presented in formats, arrangements, languages and with definitions that are primarily adapted to a specific purpose, resulting in many distinct datasets that are by default not interoperable.
However, many important downstream applications and analyses surpass the spatial, temporal or thematic scope of any unique data source or organisation, and thus rely on combining data from multiple different sources (\cite{Otto2015}).

Integrating areal data \textcolor{CornflowerBlue}{across many sources} comes with a large number of challenges (Tab.~\ref{tab:issues}) that would, if not addressed, affect database consistency and potentially bias downstream analyses.
For example, data tables that refer to the same areas must match one another, both spatially and lexically  (\cite{Du2013}), for databases to be consistent.
Besides, terms that emerge from different languages or ontologies must correspond to one another, so that a variable that has different names in different source data does in fact match (\cite{DeGiacomo2018}).
Moreover, alternative data that describe disputed areas or have been recorded by different institutions should be acknowledged to avoid bias due to erroneous political assumptions, and final outputs should be appropriately documented (\cite{Henzen2013}).

\begin{table}[!ht]

\centering
\caption{\label{tab:issues}List of issues, which have to be considered when building a database of areal data from distinct sources.}
\begin{adjustbox}{width={\textwidth},totalheight={\textheight},keepaspectratio}%
\begin{tabular}{|p{1.5in}|p{1in}|p{3in}|}
{\bf challenge} & {\bf class} & {\bf required activity} \\\hline
reproject geometries & georeferencing & harmonise spatial projection of distinct input geometries to be able to match them spatially \\
match geometries and unit names & georeferencing & connect names of territorial units to the correct geometries \\
territorial changes & alternative data & match areal data associated to territorial units that change through time \\
data sources that disagree & alternative data & match areal data associated with the same territorial units provided by different data sources \\
disputed areas & alternative data & identify data that belong to territorial units that are claimed by different authorities \\
territorial unit names & translation & translate territorial unit names into a common language \\
distinct concepts & translation & map variable names and values of categorical variables to standardised concepts \\
disorganised messy data & documentation & arrange all data in the same format \\
metadata & documentation & document dataset characteristics when input data are retrieved from the source\\
data provenance & documentation & document the procedure by which the final data product was derived \\

\end{tabular}
\end{adjustbox}

\end{table}

Several of the challenges in integrating areal data can be addressed individually with specialised tools.
For instance, a so-called ETL procedure (extract, transform, load) is typically used in data warehousing, where data from different sources are integrated into a single database (\cite{Baumer2017}, \cite{Debroy2018}).
Moreover, geometries can be matched with GIS software such as QGIS, and ontologies that describe and relate distinct concepts can be created with the software Protégé (\cite{Horridge2011}).
Finally, data can be made available via so-called Spatial Data Infrastructures (\cite{Brink2017}) \textcolor{CornflowerBlue}{and data values of the resulting database can be cleaned or validated with R-packages such as \texttt{dataMaid} (\cite{Petersen2019})}.
However, some specific challenges, such as to document input metadata or provenance or to reshape messy data (\cite{Wickham2014}) are typically solved in non-standardised ways (if at all) via custom scripts or macros that are developed for individual use-cases.
Oftentimes, important steps of database management are even carried out manually, e.g., by comparing information visually and entering the data by hand into an Excel file.

Notwithstanding the existence of many specialised tools, in practice, their application is not always trivial and may require specific expert knowledge or expensive proprietary software (\cite{Debroy2018}).
Moreover, none of the existing tools can address the full range of typical problems, so that combinations of independent tools are needed, which comes with further issues of interoperability of the specialised tools.
This complexity increases resource and time requirements for a comprehensive workflow, hindering successful implementation and thus scientific progress (\cite{Baumer2017}).
To date, a coherent and easy-to-use solution for integrating areal data that considers a large number of frequent issues and that is implemented in a robust and reproducible manner is still lacking.

Here, we introduce the R software package \texttt{arealDB} to address the overall challenge of integrating heterogeneous areal data in an easy-to-use framework.
The package offers a set of tools that are focused specifically on harmonisation and integration of areal data, thereby reducing complexity and increasing utility and confidence in data quality for downstream applications.
\texttt{arealDB} automates complicated and error-prone procedures, such as reshaping data tables, semantic matching from vocabularies or metadata and provenance documentation, with only limited required user-input.
Finally, it provides extensive unit testing, ensuring that all tools work as expected.

We exemplify the full functionality of \texttt{arealDB} by integrating two example datasets on the harvested area of soybean.
The first dataset (Brazil) is provided in Portuguese language and accompanied by specific geometries, while the second datasets (USA) is provided in our target language English and does not come with geometries but merely refers to the names of US counties.

\newpage

\section{Methods}
\label{sec:description}

\begin{figure}[ht]

\includegraphics[width=\textwidth]{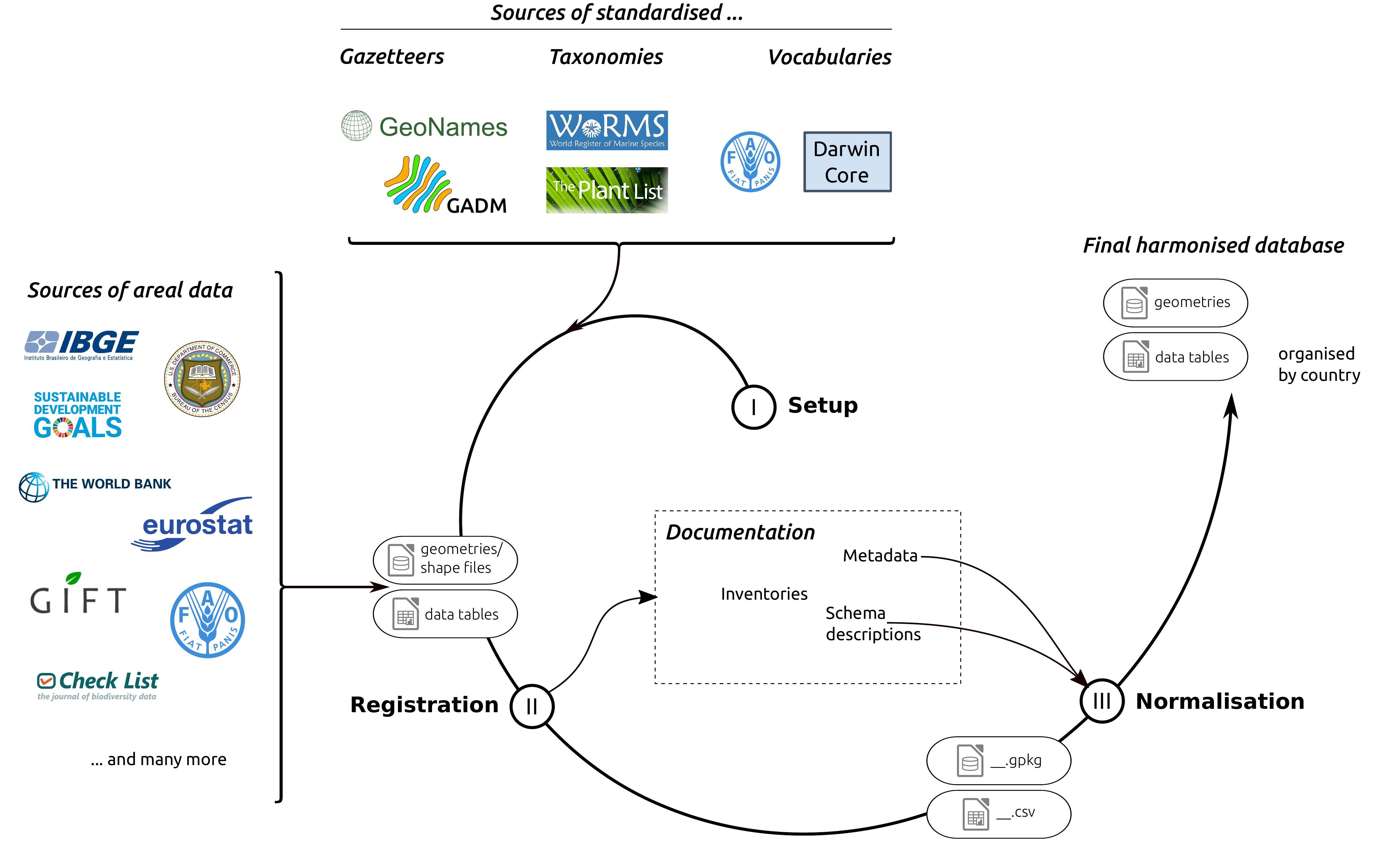}
\caption{\color{Gray} \textbf{Flow-chart of the general workflow of data integration using \texttt{arealDB}}. In stage 1, tables containing standardised terms for all variables of interest are read in to establish an ontological basis. In stage 2, data tables and geometries that have been downloaded from various data sources are registered. In stage 3, those files are harmonised and integrated into the final database. The shown data sources, gazetteers, taxonomies and vocabularies are an exemplary, non exhaustive list of sources that can be handled with \texttt{arealDB}.}
\label{fig:fig1}

\end{figure}

The R software package \texttt{arealDB} contains three groups of tools that reflect three stages of data management: (Fig.~\ref{fig:fig1}):
\begin{itemize}
	\item Stage 1 (Initialisation): Set up a database while gathering thematic metadata.
	\item Stage 2 (Registration): Transform original data from downloaded files into standardised data formats and gather metadata on those files.
	\item Stage 3 (Normalisation): Harmonise geometries and data tables based on the metadata collected at stage 2, and integrate them in a standardised database.
\end{itemize}

Technical documentation of any function that comes with this package can be retrieved after installation, for example, via the command \verb!?setPath! for the function of that name.

\subsection{Project setup}
\label{sec:setup}

\begin{figure}[!ht]

\includegraphics[width=\textwidth]{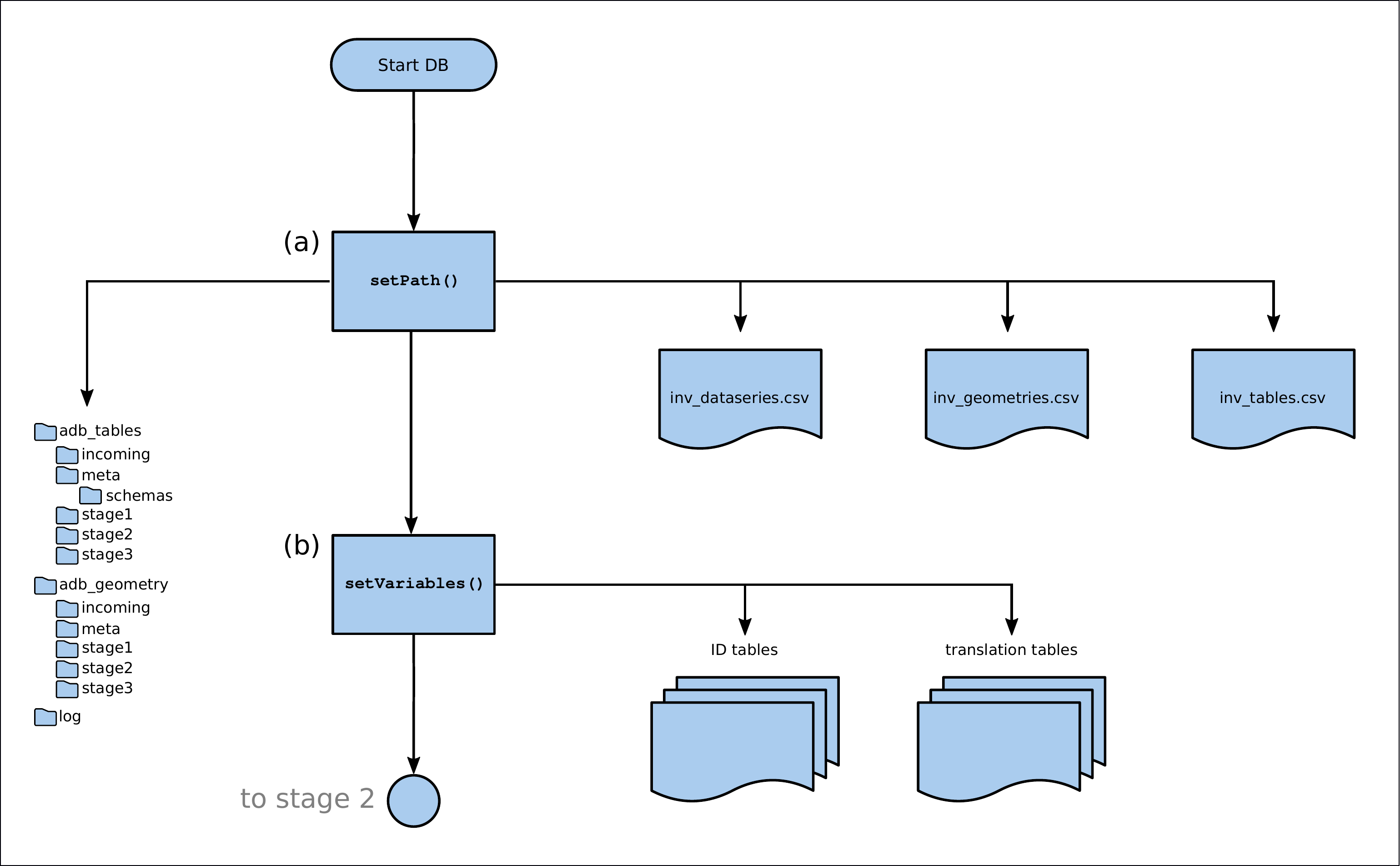}
\cprotect\caption{\color{Gray} \textbf{Flow-chart of the project setup.} \textbf{(a)} The function \verb!setPath()! initiates the project by creating a directory structure in which the files are stored and by creating the inventory tables for data-series, geometries and census tables. \textbf{(b)} The function \verb!setVariables()! creates index and translation tables for all variables that should be handled in this project.}
\label{fig:fig2}

\end{figure}

An areal database is started with the function \verb!setPath()!, which creates the standardised directory structure in which the database is stored (Fig.~\ref{fig:fig2}a).
\textcolor{CornflowerBlue}{All further operations within \texttt{arealDB} that rely on a path are then relative to this database directory.}
Any areal database typically contains a set of variables that identify the observed areal units, such as the unit names.
However, it would also include other variables that identify the observed phenomenon, such as timesteps, socioeconomic groups of people, agricultural or other commodities or biological species.
The function \verb!setVariables()! is used to setup the variables that are used in a project to handle lexical translation and semantic harmonisation of the terms of those variables (Fig.~\ref{fig:fig2}b).
It creates, by default, the skeleton of two files per variable and database, (1) an index table, which relates the variables' terms to an ID and ancillary information and (2) a translation table, which relates terms in foreign languages and semantics to the target language/ontology.
To utilize index tables, an input table that contains term-ID pairs is indispensable and translation tables do not have to, but can be provided with standardised translations of the target variables, which can help improving consistency and data quality (Fig.~\ref{fig:fig1}).

Sources of such standardised tabular information may be, for example, gazetteers such as GeoNames (\cite{geoN2019}) or the ancillary data in the Global Administrative Areas database (GADM) (\cite{Hijmans2019}), biological taxonomies such as The Plant List (\cite{List2013}), WoRMS (\cite{WoRMS20190911}) or the IUCN Red List (\cite{IUCN2019}), or standardised ontologies, such as those offered by FAOSTAT (\cite{FAO2019a}), the Darwin Core (\cite{Wieczorek2012}), the Humboldt Core (\cite{Guralnick2018}), or the Land Administration Domain Model(\cite{Lemmen2015}).

\subsection{Data registration}
\label{sec:registration}

An important aspect to ensure the quality of an integrated database is provenance documentation.
This allows tracing errors that may show only in the final database to the specific source datasets or to a certain modification process.
Documenting provenance requires that the input state of a dataset, as well as procedural metadata that become available as a side-product in the evolution from input to output data, be known.

Thus, the second stage in integrating data with \texttt{arealDB} is to create an inventory of the relevant files and to record metadata on the initial state of data, such as original file names, file locations, licenses and the arrangement of data tables (this process is called \textit{registering} in \texttt{arealDB}).
The arrangement of data tables is managed via the independent R-package \textcolor{CornflowerBlue}{\texttt{tabshiftr}} (\cite{Ehrmann2020}).
Here, so-called schema descriptions are defined, where the table-specific arrangement is described by the position (columns and rows) of data components in the table, which is the basis for automatically reshaping the files in stage 3.

\texttt{arealDB} thoroughly documents metadata on three kinds of information, data-series, geometries and data tables.
Data-series are collections of data that are provided by the same source and which share more or less the same tabular arrangement and organisational logic.
By documenting the data-series, one mostly documents information about a particular data provider and creates a tag that is common for input data that "belong together" and typically share a common structure.
That information will eventually be documented in the three inventory tables of the respective names \verb!inv_dataseries.csv!, \verb!inv_geometries.csv! and \verb!inv_tables.csv!.
None of these inventory tables ever need to be modified by hand, as all information documented here is managed automatically.

\begin{figure}[!ht]

\includegraphics[width=\textwidth]{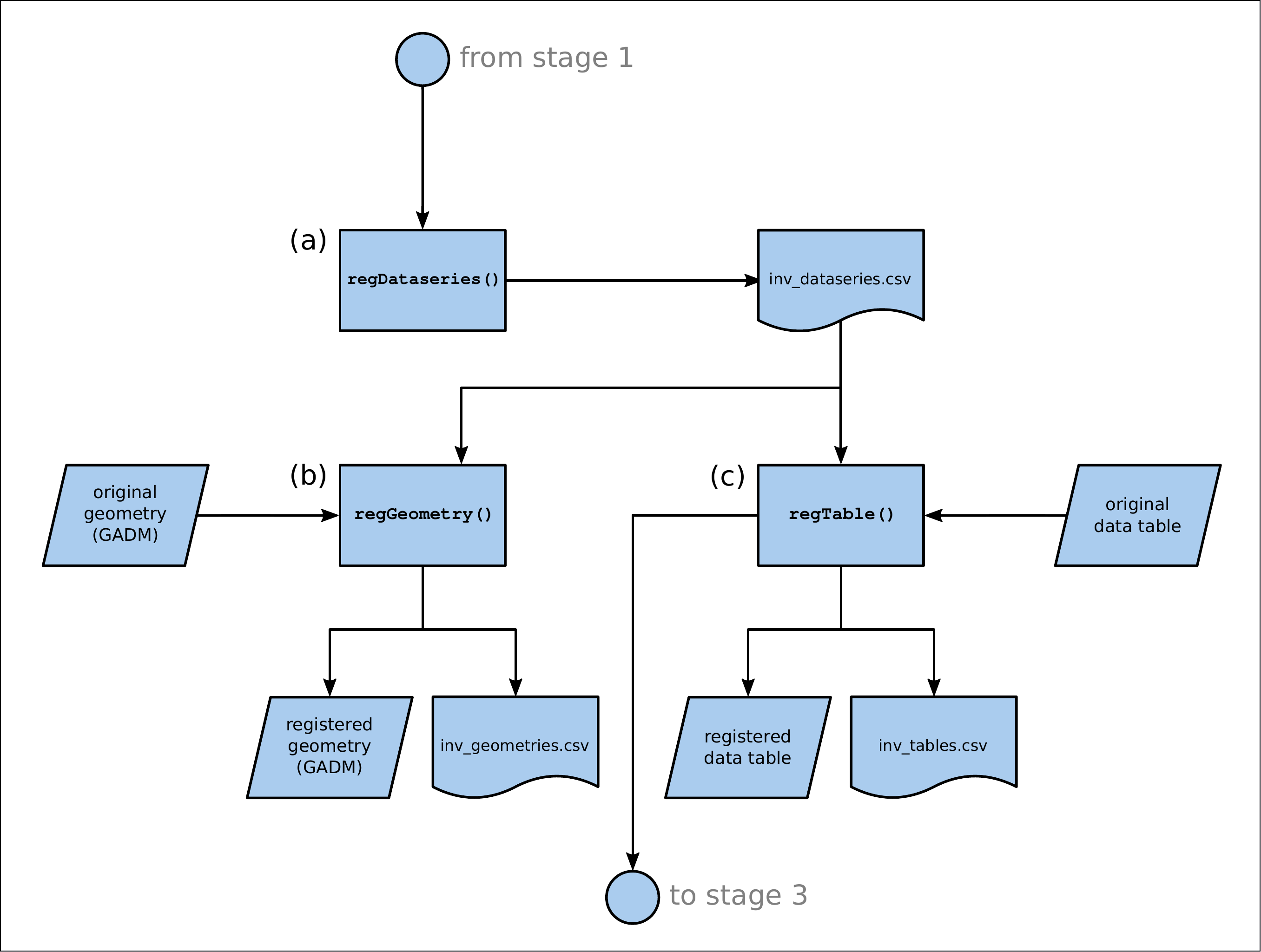}
\cprotect\caption{\color{Gray} \textbf{Flow-chart of the registration procedure.} \textbf{(a)} The function \verb!regDataseries()! is used to document the various data-series (both, of data tables and geometries) that are provided by the data source. \textbf{(b)} The function \verb!regGeometry()! is then used to register all geometry files that have been downloaded. \textbf{(c)} Finally, the function \verb!regTable()! is used to register all census tables that have been downloaded, and to relate the census tables to data-series and geometries. The registered files are stored in the folders \verb!"/adb_geometries/stage2"! and \verb!"/adb_tables/stage2"!.}
\label{fig:register_data}

\end{figure}

A new data-series is registered with the function \verb!regDataseries()!, before geometries and data tables, which are registered with the functions \verb!regGeometry()! and \verb!regTable()! (Fig.~\ref{fig:register_data}).
The functions carry out the following operations:

\begin{itemize}
	\item check the arguments for valid values and consistency,.
	\item oversee that the individual items are transformed to the target format with standard names and stored in the correct directory,
	\item create IDs for all items and insert the provided metadata into the respective inventory tables.
\end{itemize}

\subsection{Data normalisation}
\label{sec:normalisation}

\begin{figure}[!ht]

\includegraphics[width=\textwidth]{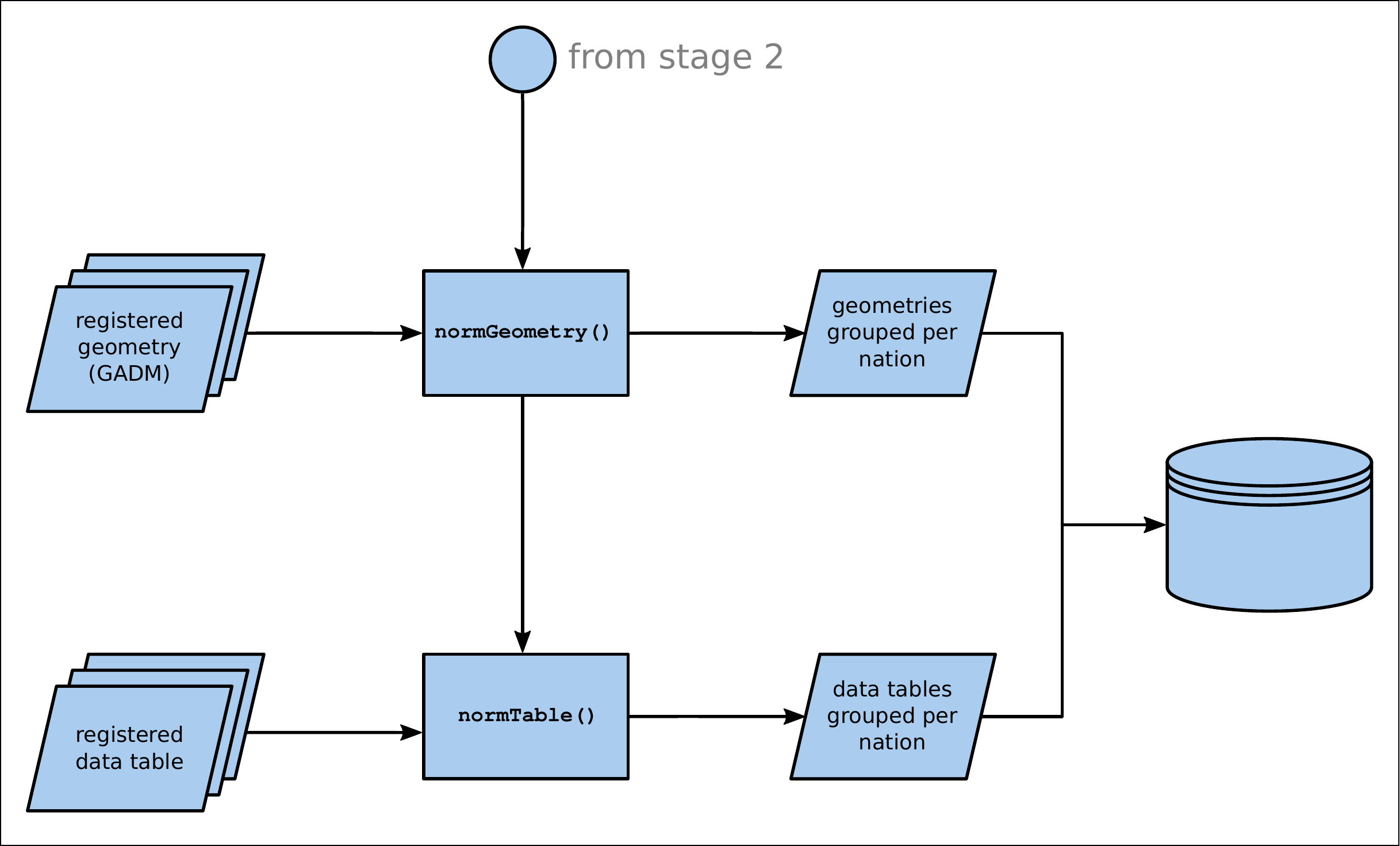}
\cprotect\caption{\color{Gray} \textbf{Flow-chart of the normalisation procedure}. The function \verb!normGeometry()! groups geometries per nation and creates the \textit{administrative hierarchy ID (ahID)}. The function \verb!normTable()! reshapes the data tables into tidy format, calls \verb!matchUnits()! to assign ahID to the areal data and groups the tables per nation. The normalised files are stored in the folders \verb!"/adb_geometries/stage3"! and \verb!"adb_tables/stage3"!}
\label{fig:normalise_data}

\end{figure}

The third and final step to integrate areal data consists of reshaping and harmonising the output of stage 2 (Fig.~\ref{fig:normalise_data}) and this process is called \textit{normalising} in \texttt{arealDB} (\cite{Codd1990}).
This final step is crucial, since at stage 2, there is still no guarantee that names of territorial units in geometries are associated with those in data tables or that areal data are georeferenced, that variables are provided in the same language across several sources, or that data tables are provided in a compatible arrangement.

Geometries are typically provided as shape or geopackage files, which have already been optimised for interoperability, and where it is thus sufficient to know which columns in the attribute table contain names of the territorial units (see \verb!regGeometry()!).
The function \verb!normGeometry()! builds harmonised geometry collections with regard to coordinate reference system, as well as semantically interoperable attribute tables.
The overall procedure is detailed in Fig. A.1.

There is no generally accepted way of recording data in tables so that these can be vastly more complex or messy than geometries, and hence require schema descriptions.
The schema descriptions that have been recorded at stage 2 document accurate positions of variables held in data tables.
The function \verb!normTable()! utilises those schemas, in concert with the function \textcolor{CornflowerBlue}{tabshiftr::reorganise()}, to reshape the data into syntactically interoperable tables.
It utilises, furthermore, the functions \verb!matchUnits()! and \verb!matchVars()! to harmonise labels of territorial units and the levels of identifying variables to end up with semantically interoperable data tables.

\subsection{The administrative hierarchy}
\label{sec:hierarchy}

Geometries in \texttt{arealDB} are assigned a unique ID for each territorial unit at each administrative level.
This requires, first of all, an initial geometry dataset from which this \textit{administrative hierarchy ID (ahID)} can be constructed, and which must thus include information on the hierarchical arrangement of the territorial units.
At each administrative level a three-digit ID is assigned to the alphabetically sorted unit names.
When descending into a lower level, that ID is restarted at 0 within each parent unit.
For instance, Tartu County in Estonia has the ahID \verb!070013!, as Estonia is the 70th country (alphabetically) and Tartu County the 13th county within Estonia.

\subsection{Translating terms}
\label{sec:translating}

When handling data from sources that span large spatial extents, these are likely only present in different languages.
However, terms may be provided not only in different languages (\textit{sensu stricto}), but also with distinct semantic meanings.
For example, the concept of \textit{patch of land that is dominated by grassy vegetation and on which cattle graze}, could be called "pasture" (British English), but also "rangeland" (American English), and is called "pastagem" in Portuguese.
The language translation "pastagem <-> pasture" from Portuguese to English is functionally similar to the semantic translation "rangeland <-> pasture".
Both examples are cases of many-to-one translations because terms in different languages (\textit{sensu lato}) refer to the same term in the target language.
Additionally, in a particular dataset, the term "pasture" might refer to anything from \textit{atificially maintained grassland managed for livestock grazing} to \textit{natural grassland} (i.e., with or without livestock). 
Ideally, these different meanings will become evident from available metadata.
In \texttt{arealDB}, an individual term is allowed to refer to different concepts, depending on where it originates, which constitutes a one-to-many semantic translation.

Many-to-one translations are handled quite straightforwardly, in that the target value is repeated in the column \verb!target! for each translation and terms that refer to it are recorded in the column \verb!origin! (Tab.~\ref{tab:tableTranslations}).
One-to-many translations are provided, in the column \verb!source!, either with \verb!geoID! or \verb!tabID!, depending on whether the terms originate from geometries or tables, and in \verb!ID! with the respective ID.

\begin{table}[!ht]

\centering
\cprotect\caption{\label{tab:tableTranslations} A translation table that includes (a) many-to-one translations (lines 1 and 2) and a case of one-to-many translations (line 3), as well as (b) language (line 1) and semantic (lines 2 and 3) translations of the term 'pasture'.}
\begin{tabular}{|l|l|l|c|c|c|c|}
{\bf origin} & {\bf target} & {\bf source} & {\bf ID} & {\bf notes} \\
\hline
pastagem & pasture &  &  & \\
rangeland & pasture &  &  & \\
pasture & grassland & tabID & 3 &  \\
... &  &  &  & \\
\end{tabular}

\end{table}

The function \verb!translateTerms()! manages all translations by comparing new terms individually and explicitly with the translation tables that have been created in stage 1.
The user is provided with an interface that suggests a range of terms pre-selected from the translation table via approximate string matching (fuzzy matching).
The missing translations then have to be provided by the user, so that the new terms can finally be compared against the look-up section of the translation table to check for consistent translations.

\section{Results}
\label{sec:results}

\begin{figure}[!ht]

\includegraphics[width=\textwidth]{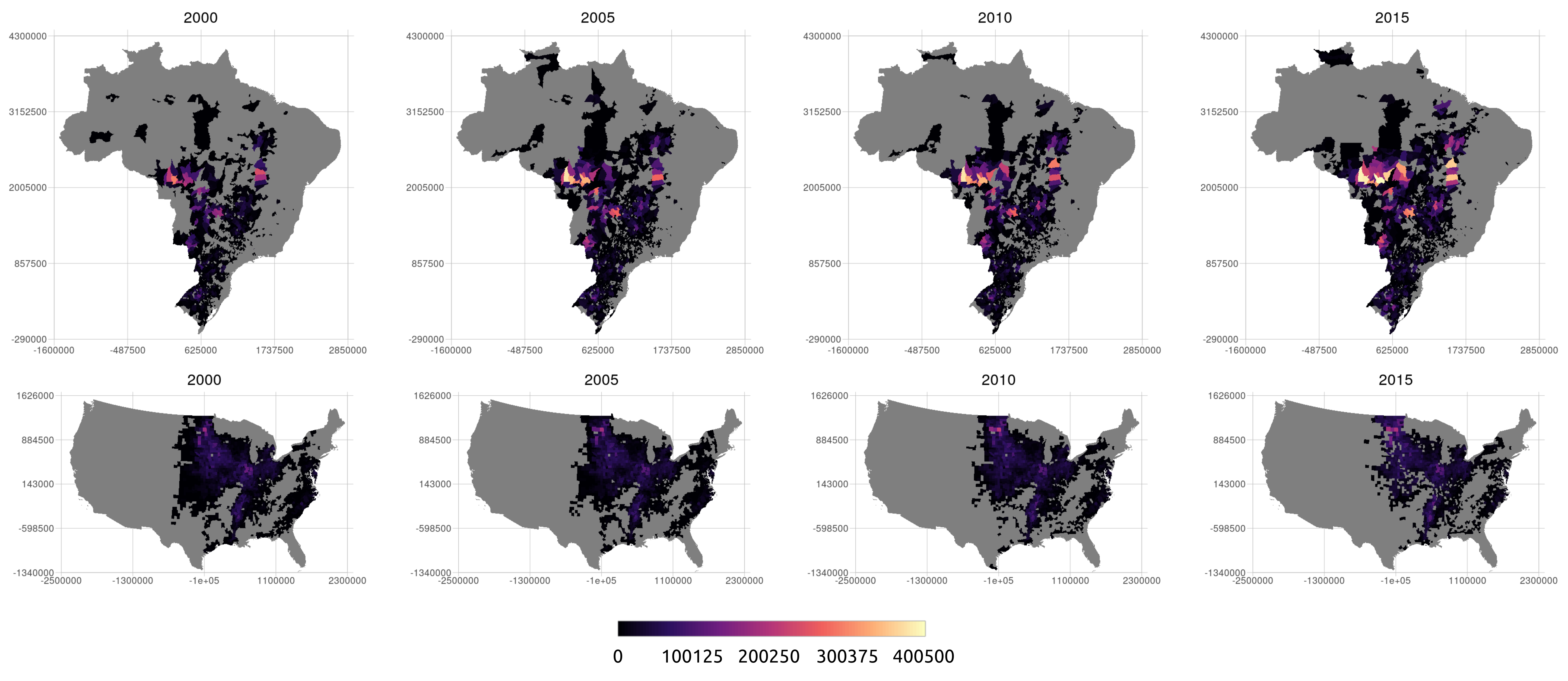}
\cprotect\caption{\color{Gray} \textbf{Choropleth maps of the final, integrated database}. The maps have been created with the simple code snippet that is shown in Appendix B. The standardised geometries and data tables were read in with a loop through all countries of interest and the maps were created with another loop through the years of interest to plot the data via the R-package \texttt{geometr} (\cite{Ehrmann2019}).}
\label{fig:output_maps}

\end{figure}

We have tested the described functionalities on several dozen agricultural and forestry census datasets from most countries of the American continent.
These data represent many of the challenges outlined in Tab.~\ref{tab:issues}, including different table arrangements of the areal data, dissimilar variables provided in different languages and associations to geometries, if they were provided at all, in distinct spatial projections. 
In Appendix A, we exemplarily showcase the integration of data on the soybean harvested area for Brazil and the US.
Downstream applications, such as maps of the spatial patterns (Fig.~\ref{fig:output_maps} and Appendix B) or explorative analysis, can profit from the relative ease of accessing all data in a database that has been built with \texttt{arealDB} at once.

The schema descriptions for each input dataset are central to this workflow.
They record the \verb!type! of each variable, the locationin a spreadsheet (\verb!row!, \verb!col!) at which the variable is and other metadata that are relevant to reshape the table.

At stage 2, the Brazilian dataset is organised in a systematic way, however the arrangement is rather complicated (Listing 1).
First of all, the spreadsheet contains a metadata header in the first three rows so that the origin of the table is at the fourth row in the first column.
The values of the identifying variables \verb!"years"! and \verb!"commodities"! and the target variable \verb!"harvested"! are stored in the same columns.
Moreover, states (\verb!al2!) and municipalities (\verb!al3!) are combined in the first column, so that this column has to be split.
The names of the states are abbreviated, requiring translation to correct names (e.g., "RO <-> Rondônia").

\begin{lstlisting}[language=R, caption=Schema description of the Brazilian dataset]
library(readr)
# input and schema for Brazil
read_csv(file = paste0(dbPath, "/adb_tables/stage2/processed/bra_3_soy_2000_2018_ibge.csv"), col_names = FALSE)
#> # A tibble: 5,569 x 20
#>    X1     X2     X3     X4    X5    X6    X7    X8    X9    X10  
#>    <chr>  <chr>  <chr>  <chr> <chr> <chr> <chr> <chr> <chr> <chr>
#> 1  Tabel  NA     NA     NA    NA    NA    NA    NA    NA    NA   
#> 2  Varia  NA     NA     NA    NA    NA    NA    NA    NA    NA   
#> 3  Munic  Ano x  NA     NA    NA    NA    NA    NA    NA    NA   
#> 4  NA     2000   2001   2002  2003  2004  2005  2006  2007  2008 
#> 5  NA     Soja   Soja   Soja  Soja  Soja  Soja  Soja  Soja  Soja
#> 6  Alta   -      -      -     -     100   -     -     -     -    
#> 7  Ariqu  -      -      450   -     -     -     50    -     -    
#> 8  Cabix  200    486    600   1500  1500  5370  7500  6000  7000 
#> 9  Cacoa  -      -      -     -     -     -     -     -     -    
#> 10 Cerej  2700   3353   3400  4516  7184  8000  18000 16200 18000
#> #   with 5,559 more rows, and 10 more variables: X11 <chr>,
#> #   X12 <chr>, X13 <chr>, X14 <chr>, X15 <chr>, X16 <chr>,
#> #   X17 <chr>, X18 <chr>, X19 <chr>, X20 <chr>

read_rds(path = paste0(dbPath, "/adb_tables/meta/schemas/schema_1.rds"))
#>   1 cluster
#>     origin: 4|1  (top|left)
#> 
#>    variable      type     row   col    rel   dist
#>   ------------- -------- ----- ------ ----- ------
#>    al2           id             1       F     F
#>    al3           id             1       F     F
#>    year          id       4     2:20    F     F
#>    commodities   id       5     2:20    F     F
#>    harvested     measured       2:20    F     F

\end{lstlisting}

The US dataset, on the other hand, is already "tidy" at stage 2, i.e., all variables are recorded in individual columns that simply have to be selected, and no lexical or ontological translations are required (Listing 2).

\begin{lstlisting}[language=R, caption=Schema description of the US dataset]
read_csv(file = paste0(dbPath, "/adb_tables/stage2/processed/usa_3_soy_2000_2018_usda.csv"), col_names = FALSE)
#> # A tibble: 30,143 x 21
#>    X1     X2    X3     X4    X5    X6    X7    X8    X9    X10  
#>    <chr>  <chr> <chr>  <chr> <chr> <chr> <chr> <chr> <chr> <chr>
#>  1 Progr  Year  Period Week  Geo   State Stat  Ag D  Ag D  Coun
#>  2 SURVEY 2018  YEAR   NA    COUN  ALAB  1     BLAC  40    DALL
#>  3 SURVEY 2018  YEAR   NA    COUN  ALAB  1     BLAC  40    ELMO
#>  4 SURVEY 2018  YEAR   NA    COUN  ALAB  1     BLAC  40    OTHE
#>  5 SURVEY 2018  YEAR   NA    COUN  ALAB  1     BLAC  40    PERRY
#>  6 SURVEY 2018  YEAR   NA    COUN  ALAB  1     BLAC  40    SUMT
#>  7 SURVEY 2018  YEAR   NA    COUN  ALAB  1     COAS  50    BALD
#>  8 SURVEY 2018  YEAR   NA    COUN  ALAB  1     COAS  50    OTHE
#>  9 SURVEY 2018  YEAR   NA    COUN  ALAB  1     MOUN  20    BLOU
#> 10 SURVEY 2018  YEAR   NA    COUN  ALAB  1     MOUN  20    CHER
#> #   with 30,133 more rows, and 11 more variables: X11 <chr>,
#> #   X12 <chr>, X13 <chr>, X14 <chr>, X15 <chr>, X16 <chr>,
#> #   X17 <chr>, X18 <chr>, X19 <chr>, X20 <chr>, X21 <chr>

read_rds(path = paste0(dbPath, "/adb_tables/meta/schemas/schema_2.rds"))
#>   1 cluster (whole spreadsheet)
#> 
#>    variable      type     row   col   rel   dist
#>   ------------- -------- ----- ----- ----- ------
#>    al2           id             6      F     F
#>    al3           id             10     F     F
#>    year          id             2      F     F
#>    commodities   id             16     F     F
#>    harvested     measured       20     F     F
\end{lstlisting}

After normalising (at stage 3), both tables share the same arrangement, where every information is encoded by IDs that point either to metadata (\verb!tabID!, \verb!geoID!), territorial units (\verb!ahID!), or to commodities (\verb!faoID!).

\begin{lstlisting}[language=R, caption=Summary of the core table of the harmonised and integrated database]
read_csv(file = paste0(dbPath, "/adb_tables/stage3/brazil.csv"), col_types = "iiiiidi") %>% 
  summary()
#>        id            tabID       geoID        ahID         
#>  Min.   :    1   Min.   :1   Min.   :5   Min.   :32001001  
#>  1st Qu.:24648   1st Qu.:1   1st Qu.:5   1st Qu.:32011029  
#>  Median :49296   Median :1   Median :5   Median :32015197  
#>  Mean   :49296   Mean   :1   Mean   :5   Mean   :32015793  
#>  3rd Qu.:73944   3rd Qu.:1   3rd Qu.:5   3rd Qu.:32021306  
#>  Max.   :98591   Max.   :1   Max.   :5   Max.   :32027137  
#>                                                            
#>       year        harvested          faoID    
#>  Min.   :2000   Min.   :     0   Min.   :236  
#>  1st Qu.:2004   1st Qu.:   460   1st Qu.:236  
#>  Median :2009   Median :  2950   Median :236  
#>  Mean   :2009   Mean   : 12388   Mean   :236  
#>  3rd Qu.:2014   3rd Qu.: 11500   3rd Qu.:236  
#>  Max.   :2018   Max.   :411224   Max.   :236  
#>                 NA's   :65529
                
read_csv(file = paste0(dbPath, "/adb_tables/stage3/united states of america.csv"), col_types = "iiiiidi") %>% 
  summary()
#>        id            tabID       geoID        ahID          
#>  Min.   :    1   Min.   :2   Min.   :3   Min.   :238001001  
#>  1st Qu.: 6871   1st Qu.:2   1st Qu.:3   1st Qu.:238016074  
#>  Median :13740   Median :2   Median :3   Median :238025054  
#>  Mean   :13740   Mean   :2   Mean   :3   Mean   :238026647  
#>  3rd Qu.:20610   3rd Qu.:2   3rd Qu.:3   3rd Qu.:238036065  
#>  Max.   :27480   Max.   :2   Max.   :3   Max.   :238050074  
#>       year        harvested             faoID    
#>  Min.   :2000   Min.   :    40.47   Min.   :236  
#>  1st Qu.:2004   1st Qu.:  2954.21   1st Qu.:236  
#>  Median :2008   Median : 12788.07   Median :236  
#>  Mean   :2009   Mean   : 20302.90   Mean   :236  
#>  3rd Qu.:2013   3rd Qu.: 33108.34   3rd Qu.:236  
#>  Max.   :2018   Max.   :218125.56   Max.   :236
\end{lstlisting}

\section{Discussion}
\label{sec:discussion}

\begin{figure}[!ht]

\includegraphics[width=\textwidth]{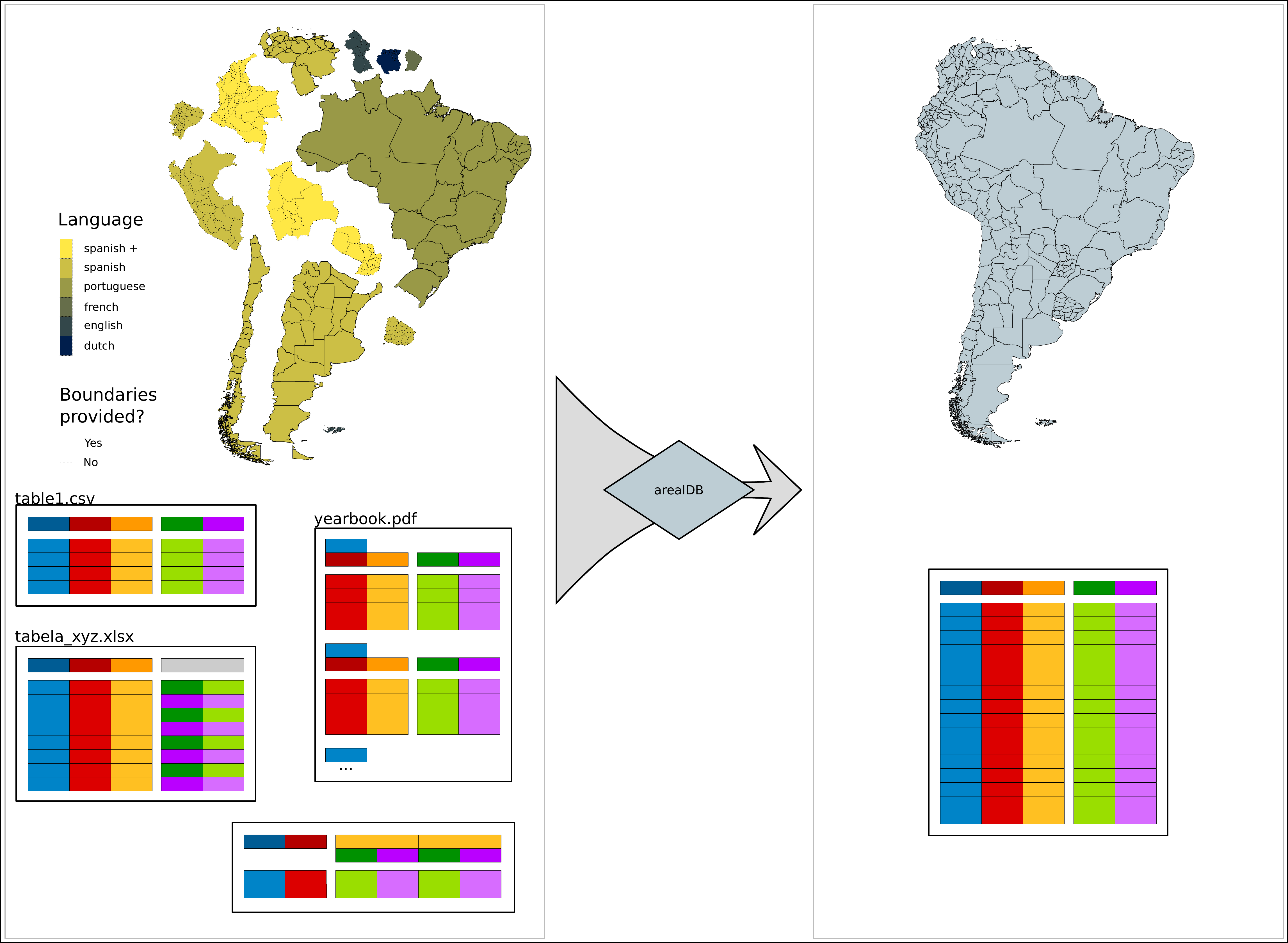}
\cprotect\caption{\color{Gray} \textbf{Schematic overview of the purpose of \texttt{arealDB}}. Both, data tables that are disorganised and messy and that are based on different languages and concepts, as well as distinctly organised geometric datasets are harmonised and integrated into one consistent database.}
\label{fig:schematic_overview}

\end{figure}

The present R package \texttt{arealDB} provides so far missing software for harmonising and integrating areal data across multiple heterogeneous sources into a single, consistent database (Fig.~\ref{fig:schematic_overview}).
By guiding users through the three stages 'project setup', 'data registration' and 'data normalisation' , embedded into a collaborative and transparent software environment (\cite{Lowndes2017}), \texttt{arealDB} substantially improves the speed, scientific accuracy, and reproducibility \textcolor{CornflowerBlue}{of integrating areal data}.

Lacking such a tool, many data integrators have thus far gone through a highly time-consuming and error-prone process of opening individual input datasets in Excel, translating variables "by hand", and reshaping tables by copying relevant data from one spreadsheet into another, typically without any provenance documentation.
More sophisticated workflows based on advanced tools, such as lookup and pivot tables or programmable statistical/data management tools often require custom-scripted solutions for the different, heterogeneous input data-sets.
\texttt{arealDB}, by contrast, merely requires that users document metadata, provide translation tables and specify a schema description for each dataset.

\subsection{Handling geometries}
\label{sec:geometry_handling}

Using \texttt{arealDB}, each individual areal data table may be linked to a different geometry dataset.
This helps to avoid political and other assumptions where more than one source of areal data exists for territorial units, for example in cases of disputed areas or administrative changes.
Data that refer to such disparate geometry sources can coexist in a database without biasing downstream analyses that are sensitive to the areas of measurement units (e.g., when estimating ecological scaling relationships from species checklists; (\cite{Kreft2007}; \cite{Keil2019}).
Where such considerations are not an issue, datasets may be linked to standardised geometries, such as GADM (\cite{Hijmans2019}).

Currently, \texttt{arealDB} matches areal data via an internal assessment of the geographical overlap of geometries (using the R package \texttt{sf} (\cite{Pebesma2018}); see flowchart in Appendix C).
This procedure could be further refined by additionally incorporating functions that can automatically detect and handle temporal changes of territorial units (\cite{Bernard2018}). 
However, as all raw input geometries are retained in the final database, more sophisticated procedures for matching recorded geometries may be applied alongside \texttt{arealDB}'s own functions.

Another common challenge in matching territorial units lies in names that are shared by multiple territorial units located in different nations and at different administrative levels.
For example, the term 'Santa Cruz' may refer to city districts, municipalities, departments, or other units in over twenty different nations.
\texttt{arealDB} addresses this issue by matching territorial unit names to geometries hierarchically.

\subsection{Documenting metadata}
\label{sec:metadata}

Even \texttt{arealDB} is still prone to some human error, mostly related to correctly reading in the required files and information.
\texttt{arealDB}'s system of unique IDs ensures that all data values can be linked to metadata on their input tables, associated geometries, and original data sources.
Hence, inconsistencies that may still exist in an output database are fully traceable, due to the metadata collected while collating the database (\cite{Henzen2013}).
Moreover, the relational setup of the resulting database allows that database management software can retrieve metadata on all harmonised datasets (e.g., the recorded number of entities or observations, or value ranges) and make those and other metadata available also to other tools and environments.

\subsection{Interoperability}
\label{sec:interoperability}

\texttt{arealDB} was primarily designed for the purpose of integrating heterogeneous areal data within a given knowledge domain.
However, by storing all variables in the same data structure, the tools provided here enable areal databases that are syntactically interoperable across domains.
The areal data in each and every output table of \texttt{arealDB} have a value of ahID (Code-Listing 3) that is by default always derived from the same spatial basis, the GADM dataset. 
Merely areal data for which ahID was derived from specific non-GADM geometries may deviate from a common list of ahID values.
This means that areal databases from several distinct domains (for instance, human health and biodiversity) that have all been built with the same geometries dataset can simply be joined via ahID to combine information on the distinct topics within a consistent database and facilitate interdisciplinary applications (\cite{Otto2015}).

While \texttt{arealDB} enables such data integration technically, the endeavour of actually integrating databases across knowledge domains in a meaningful way hinges crucially on advances in ontological standardisation to come up with concepts that are valid simultaneously across knowledge domains (\cite{Brink2017}).

\section{Conclusions}
\label{sec:conclusions}

\texttt{arealDB} provides a range of tools that standardise the process of integrating heterogeneous areal data sources into a single, harmonised database, removing hurdles that come with high resource and time requirements.
It enables users that may lack the necessary background knowledge to address the majority of the anticipated issues in setting up a coherent workflow of data integration.
This helps in homogenising data collection methodologies, which is urgently needed to tackle data management strategies that are able to deal with vast amounts of heterogeneous data (\cite{Otto2015}).

The tools presented here allow to process any areal data, such as socioeconomic census and survey data, ecological checklist data, data on infectious diseases, (sub-)national indicator data, cadastral parcel data, and many more.
\texttt{arealDB} lowers the effort to surpass barriers of the spatial, temporal or thematic scope of single data sources.
This can enable many exciting and easy to set-up applications at the pan-regional and global scale, supporting data-integration needs across application domains and progressing towards multiple Sustainable Development Goals.

\section{Acknowledgements}
\label{sec:acknowledgements}
CM acknowledges funding by the Volkswagen Foundation via a Freigeist Fellowship. SE acknowledges funding by iDiv via the Flexpool mechanism (FZT-118, DFG).

\section{Appendix}
\label{sec:appendix}

\begin{itemize}
	\item Appendix A: Replication Script
	\item Appendix B: Downstream Application
	\item Appendix C: Normalise Geometries
\end{itemize}

\bibliography{manuscript}
\bibliographystyle{plainnat}

\end{document}